# Crystal structure, superconductivity and magnetic properties of the superconducting ferromagnets $Gd_{1.4-x}Dy_xCe_{0.6}Sr_2RuCu_2O_{10}$ (x = 0 - 0.6)

S. Kalavathi, J. Janaki, G.V.N. Rao[*], V. Sankara Sastry and Y. Hariharan

Materials Science Division, Indira Gandhi Centre for Atomic Research, Kalpakkam 603 102, Tamil Nadu, India

*International Advanced Research Centre for Powder Metallurgy & New Materials, Balapur, Hyderabad 500 005, Andhra Pradesh, India

---------------------------------------------------------------------------------------------------------------

## Abstract

The structural, electrical and magnetic properties of the superconducting ferromagnets, $Gd_{1.4-x}Dy_xCe_{0.6}Sr_2RuCu_2O_{10}$ (x=0-0.6) are systematically investigated as a function of Dy doping and temperature. These compounds are characterized by high temperature superconductivity ( $T_c$ ranging from 20-40 K depending upon the Dy content ) coexisting with weak ferromagnetism with two magnetic transitions ($T_{M2}$ ranging from 95-106 K and $T_{M1}$ around 120K). Doping with Dy gives no significant structural changes except for a minor change in the c/a ratio. However the superconducting transition temperature is significantly suppressed and magnetic ordering temperature enhanced on Dy doping. These effects are described and discussed.

Corresponding Author:

Ms. S. Kalavathi                                Ph: 04114 – 480081

Scientific Officer                              e- mail: kala@igcar.ernet.in

Materials Science Division

Indira Gandhi Centre for Atomic Research

Kalpakkam – 603 102

Tamil Nadu

India

Introduction:

Co-existence of superconductivity (SC) and magnetism is a phenomenon that has long attracted attention [1]. Materials like RERh$_4$B$_4$ ( RE= Nd, Sm and Tm) and REMo$_6$S$_8$ (RE= Gd, Tb, Dy and Er) have shown co-existence of superconductivity and antiferro magnetism (AFM). In high Tc superconducting oxides the ordering of the rare earth moments at very low temperature do not have any effect on the superconductivity which occurs in the copper oxide layers with appropriate charge transfer from adjacent reservoirs. More intriguing is the co-existence of ferromagnetism (FM) and superconductivity. ErRh$_4$B$_4$ and HoMo$_6$S$_8$ show such a coexistence of FM and SC in a very narrow window of temperature. In the case of some of the borocarbide superconductors the scenario is more complex [2]. For example the compound ErNi$_2$B$_2$C is superconducting below 10.5 K and orders antiferromagnetically below 6.0 K with a fundamental incommensurate wave vector of (0.553π/a, 0, 0) where a is the lattice constant in the plane of this tetragonal crystal. But at temperatures below 2.3 K it orders ferromagnetically with a moment of roughly 0.33 $\mu_B$ / Er. Many interesting consequences can arise when such a co-existence occurs like coexistence of triplet superconductivity and ferromagnetism [3] that occurs in UGe$_2$ , ZrZn$_2$ etc. which are itinerant ferromagnets. In these systems the same electrons that contribute to FM contributes to SC and when FM disappears superconductivity too disappears. Other consequences can be the formation of spontaneous vortex phase [4], spatially inhomogeneous Fulde-Ferrel-Larkin-Ovchinnikov [5] type superconducting order etc. Thus the observation [6] of superconductivity coexisting with weak ferromagnetism in the hybrid rutheno-cuprate systems $R_{1.4}Ce_{0.6}Sr_2RuCu_2O_{10}$ and $RSr_2RuCu_2O_8$ (R=Sm, Eu, Gd) has triggered intense research activity in this field both with respect to synthesis and physico-chemical characterization of new compositions prepared by substitution and studying the nature of superconductivity, magnetism and their inter-relationship.

The structure of the rutheno-cuprates is closely related to the well known cuprate superconductor $YBa_2Cu_3O_{7-\delta}$. Fig. 1 shows the idealized version of the structure of $Gd_{1.4}Ce_{0.6}Sr_2RuCu_2O_{10}$. It can be derived from $YBa_2Cu_3O_7$ by the replacement of $CuO_2$ chains with $RuO_6$ octahedra , Ba by Sr and the insertion of a fluorite type $(Ln, Ce)O_2$ layer between the bases of the $CuO_5$ pyramids. The latter shifts the alternate perovskite blocks by $(\underline{a}+\underline{b})/2$ leading to a body centered unit cell as shown in the figure with space group I4/mmm.

Several studies have been reported in literature on $Gd_{1.4}Ce_{0.6}Sr_2RuCu_2O_{10-\delta}$ including thermal, transport, magnetic and spectroscopic properties as well as magneto optical imaging [6][7][8][9]. However the precise nature of the magnetic order has not yet been established. Specific heat studies as a function of temperature in magnetic fields from 0-6 Tesla for $Gd_{1.4}Ce_{0.6}Sr_2RuCu_2O_{10}$ by Chen et al [7] indicate clearly a sizable specific heat jump at 38 K corresponding to the superconducting transition confirming the presence of bulk superconductivity. Magnetization studies on $R_{1.4}Ce_{0.6}Sr_2RuCu_2O_{10}$ (R=Eu, Gd) by Felner et al [6] indicate that the magnetization in these samples is composed of three contributions (a) a negative moment below $T_c$ due to the superconducting state (b) a positive moment due to the paramagnetic effective moment of R ($P_{eff}$=7.94 $\mu_B$ for R=Gd) and (c) a contribution from the ferromagnetic like behavior of the Ru sub-lattice. They observe two magnetic transitions characteristic of these samples and claim that the smaller peak which occurs at higher temperature arises as a result of an anti-symmetric exchange coupling of Dzyaloshinsky-Moriya (DM) type between neighbouring moments , induced by a local distortion that breaks the symmetry of the $RuO_6$ octahedra. Due to this DM interaction the field causes the spins to cant slightly out of the original direction and to align a component of the moments with the direction of the applied field H. At low temperatures, the Ru and / or Gd-Ru interactions begin to dominate leading to the total alignment of the Ru moments and to the dominant peak at around 100 K. In the case of $Eu_{1.6}Ce_{0.4}Sr_2RuCu_2O_{10}$ they observe a small ferromagnetic hysteresis loop at 5 K below 1.5 kOe indicating weak ferromagnetism in these samples. Based on X-ray photoemission of core levels, valence band studies using angle - integrated photoemission, and magnetization measurements on ruthenocuprates including $Eu_{1.4}Ce_{0.6}Sr_2RuCu_2O_{10}$, Frazer et al [8] have concluded that the magnetic ordering occurs in the $RuO_2$ planes. The scanning–tunneling spectroscopy measurements of Felner et. al. [6] on $R_{1.4}Ce_{0.6}Sr_2RuCu_2O_{10}$ (R=Eu,Gd) reveal a superconducting gap structure for the whole sample indicating that the materials are single phase , which simultaneously manifest both superconductivity and magnetism. Magneto-optical imaging of this system [9] indicates co-existence of magnetism and superconductivity within the experimental spatial resolution of ~10 μm . A recent XANES study indicates a fixed valency of 5 for Ru [10] in the $Gd_{1.4}Ce_{0.6}Sr_2RuCu_2O_{10}$ system as against the variable valency of Ru in the $GdSr_2RuCu_2O_8$ system. The $GdSr_2RuCu_2O_8$ system (which is another hybrid rutheno-cuprate system) is very different with one single magnetic peak at about 130 K and a superconducting transition temperature at 46 K. While it is not yet conclusively proven whether the magnetic ordering in this system is the G-type AFM ordering with a possible canting of Ru moments to accommodate ferromagnetism as suggested by the analysis of neutron diffraction data [11] or an I-type ordering with intra-layer ferromagnetic ordering and interlayer anti-ferromagnetism as suggested by the analysis of magnetization data and band structure calculations [12,13], the existence of variable valency supposedly entails several ordering possibilities.

Substitution and size effects are known to affect $T_c$ in ruthenocuprates. In the case of the $RSr_2RuCu_2O_8$ system there is an increase in $T_c$ with reducing ionic size viz. $RSr_2RuCu_2O_8$ where R =Eu shows a $T_c$ of 35 K, R=Gd shows a $T_c$ of 48 K and partial replacement of Gd by Dy increases the $T_c$ to 58 K [14]. In this background it was considered interesting to study the role of Dy substitution in modifying the magnetic and superconducting properties of the $Gd_{1.4}Ce_{0.6}Sr_2RuCu_2O_{10}$ system. In order to gain further insight into the interplay of magnetism and superconductivity we have synthesized a series of doped $Gd_{1.4-x}Dy_xSr_2RuCu_2O_{10}$ compounds with Dy concentrations ranging from $0 < x < 0.6$. The samples have been characterized using X-ray diffraction, SEM-EDS analysis and investigated using susceptibility and electrical transport measurements, and the effect of partial substitution of Dy for Gd on both $T_m$ and $T_c$ is reported

Experimental:

Single phase polycrystalline samples of $Gd_{1.4-x}Dy_xCe_{0.6}Sr_2RuCu_2O_{10-\delta}$ (x=0,0.2,0.4,0.6) were synthesized by the high temperature solid state reaction technique. Stoichiometric amounts of $Gd_2O_3$, $Dy_2O_3$, $CeO_2$, $CuO$ and Ru powders, all greater than 99.9% pure, have been mixed and ground. The mixed powder was heated at 600°C for 48 hrs. to avoid $RuO_2$ volatility and then at 950-975°C in air for 12 hrs. It was then ground and pelletised and the pellets were reacted at 1050°C in Oxygen for 72 hrs. Repeated grinding and heating at 1050°C in Oxygen was carried out to obtain single phase samples. A final annealing was done at 1090°C for 72 hrs. A sample of $Eu_{1.5}Ce_{0.5}Sr_2RuCu_2O_{10}$ was also synthesized for comparison following the same sequence of synthesis steps mentioned as above. High resolution X-ray powder diffraction was carried out using a STOE diffractometer in the step scanning mode. Si powder was used as an internal standard for calibration. The STOE refinement program using a least squares procedure was used to refine the lattice constants and to calculate standard deviations for these parameters. Metallographic examination and chemical analysis have been carried out by SEM-EDS for two representative samples including $Gd_{1.4-x}Dy_xSr_2RuCu_2O_{10-\delta}$ (x=0 and 0.4) using a JEOL model JSM 5410 SEM.

Four probe electrical resistance was measured in the Van der Pauw geometry using a highly stable constant current source and a Keithley model 181 Nano voltmeter. Magnetic susceptibility signals were traced using a relative ac susceptibility coil assembly having a primary and two secondary coils wound opposite to each other. A PAR model 5210 lock-in amplifier was used to detect the signals. Both resistance and susceptibility measurements were carried out in a dip stick arrangement exploiting the temperature gradient available in the He gas above the liquid level. The samples were mounted on a Cu block at the tip of a thin walled stainless steel dipstick. In the case of resistance measurement electrical leads were taken using a silver paste contact. In the case of susceptibility measurement powdered samples were filled in a Teflon capsule and placed in the core of one of the secondary coils. The experiments were done slowly in about 5 hrs. time for tracing the signal in the temperature range from 300 K to 4.2 K for each of the samples and measurements were repeated many times to ensure reproducibility. A calibrated Si diode sensor was used to measure the temperature.

Results:

Fig. (2) shows the powder X-ray diffraction patterns of the compounds $Gd_{1.4-x}Dy_xCe_{0.6}Sr_2RuCu_2O_{10}$ for x= 0, 0.2, 0.4 and 0.6. Analysis of the XRD data was done on the basis of the reported tetragonal structure (Fig. 1) with space group I4/mmm [15]. It has been found out from the analysis that the impurity content in all the samples is less than 3% except for the composition x= 0.6 which contains substantial amount (around 15%) of the double Perovskite phase of the type $Sr_2GdRuO_6$. Table (1) lists the lattice constants of all the samples including that of $Eu_{1.5}Ce_{0.5}Sr_2RuCu_2O_{10}$. It is seen that there is an increase in both the a and c- parameters on going from $Gd_{1.4}Ce_{0.6}Sr_2RuCu_2O_{10}$ to $Eu_{1.5}Ce_{0.5}Sr_2RuCu_2O_{10}$. This is consistent with the larger average ionic radius of the rare earth ion in the latter compound relative to the former. On Dy substitution for Gd in $Gd_{1.4}Ce_{0.6}Sr_2RuCu_2O_{10}$ there is a decreasing trend in c-lattice parameter with increase in Dy content while the a- lattice parameter remains constant. This is consistent with the fact that Dy has entered the lattice. XRD of the pellets (not shown) indicate that they exhibit preferred orientation with especially increased intensity for (0 0 14) reflection among the (0 0 l) reflections. As the (0 0 14) plane corresponds to the Sr-O layer, this may be due to defects in the Sr-O layer.

SEM pictures of $Gd_{1.4-x}Dy_xCe_{0.6}Sr_2RuCu_2O_{10}$ for x = 0 and 0.4 are shown in fig. (3). The grain sizes for the composition x=0.4 are very large and greater than 10 μm whereas for the composition x = 0 they are very small (less than 5μm). The characteristic feature observed from SEM micrographs is that the aspect ratio is larger for Dy=0.4 composition as compared to Dy=0. To determine the homogeneity of chemical composition of the samples the composition of five different grains chosen by representative shades from the SEM were measured by EDS. The standard deviations of EDS intensities were calculated. The standard deviations range from 0.6 for Cu and Ce to 1.6 for Dy in the case of sample with Dy = 0.4 . Similarly the standard deviations of EDS intensities were calculated to range 1.3 for Ru to 3.5 for Cu in the case of sample with Dy =0. In this context it may be mentioned that anti site substitution namely Gd for Sr and Ru for Cu are possible in this system [16]. This may give rise to sample inhomogeneity which can reflect as standard deviation. It is to be noted that even though the same sequence of operations during the synthesis process was adopted for all samples, Dy=0.4 sample has formed with better homogeneity, larger grain size and aspect ratio.

The resistance behaviour of the samples is plotted in fig. (4). Dy=0 and Dy=0.2 the samples show an increase in resistance as the temperature is decreased till the onset of superconductivity. Whereas the Dy=0.4 and Dy=0.6 sample indicate a decrease in R as T decreases. The onset of fall in resistance signal is identified as the onset of SC and the temperature corresponding to that as the superconducting transition temperature $T_c$. It is also observed from fig. (4) that close to $T_c$ all the samples show a semiconductor like behaviour probably due to the granular nature of the sample or due to a modification in the magnetic scattering. The onset $T_c$ for Dy=0 sample ($Gd_{1.4}Ce_{0.6}Sr_2RuCu_2O_{10}$ ) is 40 K and compares well with the literature value [17]. It decreases to 32 K, 23 K and 22 K with 0.2, 0.4 and 0.6 Dy respectively as shown in the inset of fig. (4). The width of the resistive transition is about 10 – 20 K. The existence of superconductivity is verified by checking the current dependence in resistance measurement for all the samples at 4.2 K. Up to about 15 mA current across the sample pellet the superconductivity is retained and beyond that current the normal signal reappears.

The results of the a. c. susceptibility measurements are presented in fig. (5). From the figure it is seen that in the case of $Gd_{1.4}Ce_{0.6}Sr_2RuCu_2O_{10}$ there are two peaks corresponding to magnetic ordering, a small one at higher temperature of the order of 120 K and a dominant one around 95 K. The smaller magnetic peak occurring at the high temperature ($T_{M1}$) shown in fig. (6) is identified in the literature as being due to the possible anti-symmetric exchange coupling of the Dzyaloshinsky- Moria type (DM) between neighbouring Ru moments induced by a local distortion that breaks the tetragonal symmetry of the $RuO_6$ octahedra. Due to this interaction the field causes the adjacent spins to cant slightly out of their original direction thereby lending a component in the direction of the applied field. It is seen from the figure that substitution of Dy has an effect either directly or indirectly on this interaction and there by shifts $T_{M1}$. As the temperature is lowered the Ru-Ru interaction begins to dominate and this probably gives rise to the stronger susceptibility peak at $T_{M2}$. Further lowering of temperature results in a fall in the measured secondary e.m.f that does not quite become negative. Whereas the same coil assembly usually measures huge negative signals corresponding to the superconducting transition temperature of Pb, Nb, High Tc superconductors, borocarbides and recently the MgB2 powders synthesized and measured in our laboratory. Nevertheless in the present case the onset of this fall is identified as the onset of the superconductivity since it points to a change in magnetic behaviour of the sample and it is around this temperature, the fall in resistivity was also observed. The dominant peak in the susceptibility due to Ru-Ru interaction at $T_{M1}$ is also shifted to higher temperature namely from ~95 K for Dy=0 sample to ~106 K for Dy=0.6 sample. The onset of superconductivity is at 34.4 K for $Gd_{1.4}Ce_{0.6}Sr_2RuCu_2O_{10}$ and it shifts to 28.5 K and 19.2 K respectively for Dy 0.2 and Dy 0.4. When the Dy content is increased to 0.6 no fall could be observed.

Discussion:

It is observed from the present study that the substitution of Dy has certainly affected both the magnetic and super conducting transitions in the Ru-2212 system. In the absence of any other experiments to elucidate the nature of the magnetic ordering in this system we can only infer from our data that the strength of magnetic interaction may have increased on substitution of Dy. Sonin et al. [ 4 ] have proposed in their paper a possible domain structure in this system to be co-existing with superconductivity. It is also to be borne in mind that $Dy^{3+}$ ion has higher magnetic moment compared to that of $Gd^{3+}$ ion. This, in an indirect manner, might affect the canting of $Ru^{5+}$ ions in the RuO planes. Thus if the sample has been existing in a spontaneous vortex phase a further increase in internal magnetic field could very well decrease the superconducting transition temperature. This could again be the reason for the other observation in the present study viz. the $T_c$ observed in a susceptibility experiment has consistently been lesser compared to that measured by resistive experiment in the same sample. Further experiments to estimate the local magnetic field strength, the upper and lower critical fields and to reveal the nature of magnetic ordering would aid in understanding the system.

Figure Captions:

Fig. 1. Idealised Crystal structure of (Gd/ Eu)$_{1.+x}$Ce$_{1-x}$ Sr$_2$RuCu$_2$O$_{10}$

Fig. 2 X -ray diffraction patterns of Gd$_{1.4}$Ce$_{0.6}$Sr$_2$RuCu$_2$O$_{10}$ and Gd substituted with 0.2 , 0.4 and 0.6 Dy . X- ray diffraction pattern of Eu$_{1.5}$Ce$_{0.5}$ Sr$_2$RuCu$_2$O$_{10}$ is also shown for comparison. All the lines are indexed to tetragonal I4/mmm structure.

Fig.3 Grain sizes obtained from SEM corresponding to a) Gd$_{1.4}$Ce$_{0.6}$Sr$_2$RuCu$_2$O$_{10}$ and b) Gd$_{1.0}$Dy$_{0.4}$Ce$_{0.6}$Sr$_2$RuCu$_2$O$_{10}$

Fig.4 Four probe resistance as a function of temperature. The superconducting transition temperature decreases as the Dy concentration increases. Inset shows the decrease of T$_c$ as the Dy content increases

Fig.5 Variation of a.c. susceptibility as a function of temperature. A small peak appears around 120 K (T$_{M1}$) , a major magnetic peak which shits to higher temerature on Dy substitution appears at around 95 K (T$_{M2}$ ) and the onset of the fall at lower temperature indicates superconducting transition (T$_c$) . Inset shows the decrease of T$_c$ as the Dy content increases

Fig.6 Dy substitution affects the smaller magnetic peak (T$_{M1}$) due to the canting of the Ru moments brought about by the DM interaction.

| Sample | a (Å) | c (Å) | $T_c$ (K) (Resistive) |
|---|---|---|---|
| $Gd_{1.4}Ce_{0.6}Sr_2RuCu_2O_{10}$ | 3.8351(0.0008) | 28.5793(0.0070) | 40 |
| $Gd_{1.2}Dy_{0.2}Ce_{0.6}Sr_2RuCu_2O_{10}$ | 3.8355(0.0004) | 28.5833(0.0043) | 32 |
| $Gd_{1.0}Dy_{0.4}Ce_{0.6}Sr_2RuCu_2O_{10}$ | 3.8353(0.0007) | 28.5550(0.0075) | 23 |
| $Gd_{0.8}Dy_{0.6}Ce_{0.6}Sr_2RuCu_2O_{10}$ | 3.8340(0.0006) | 28.5472(0.0055) | 22 |
| $Eu_{1.5}Ce_{0.5}Sr_2RuCu_2O_{10}$ | 3.8415(0.0006) | 28.5919(0.0057) | 40 |

Table 1. Lattice parameter and Resistively measured transition temperatures

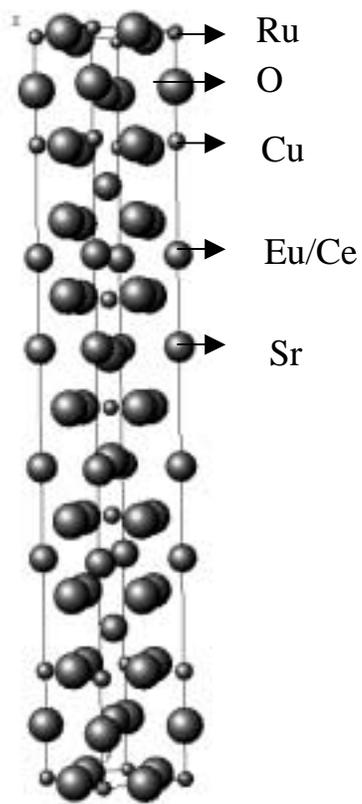

Fig. 1

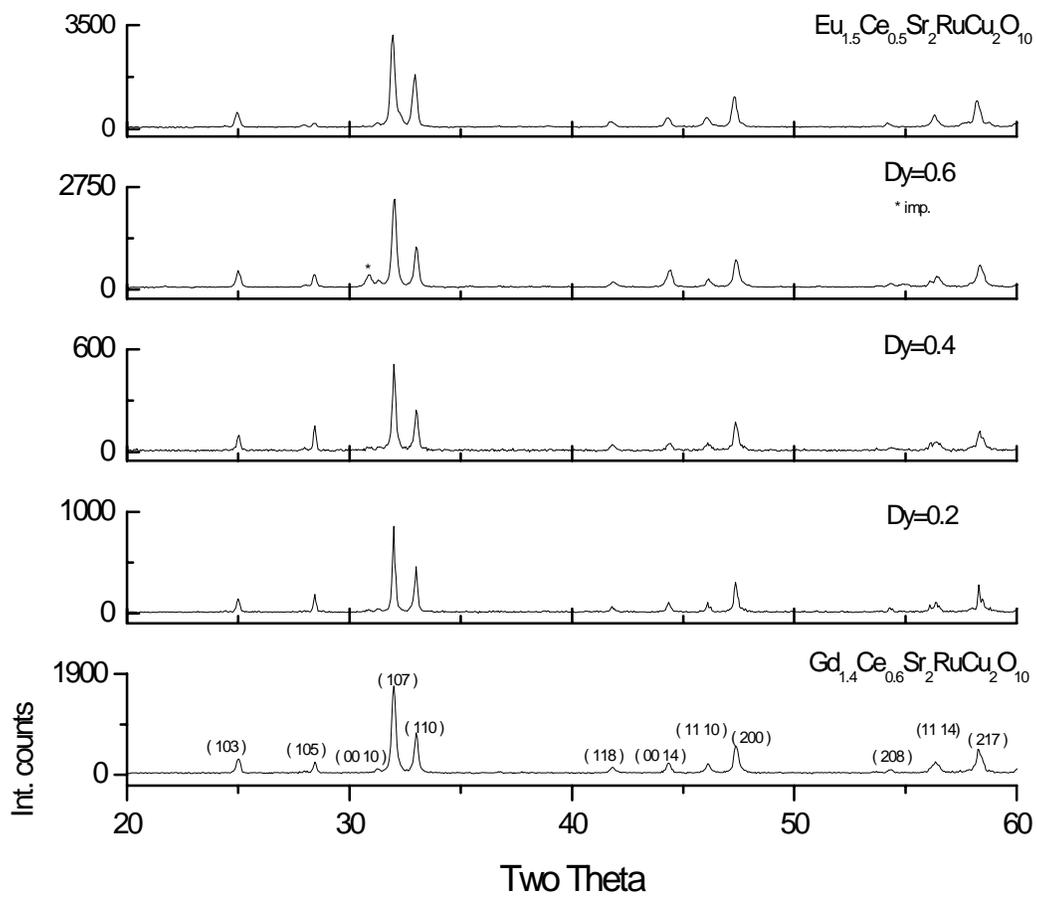

Fig. 2

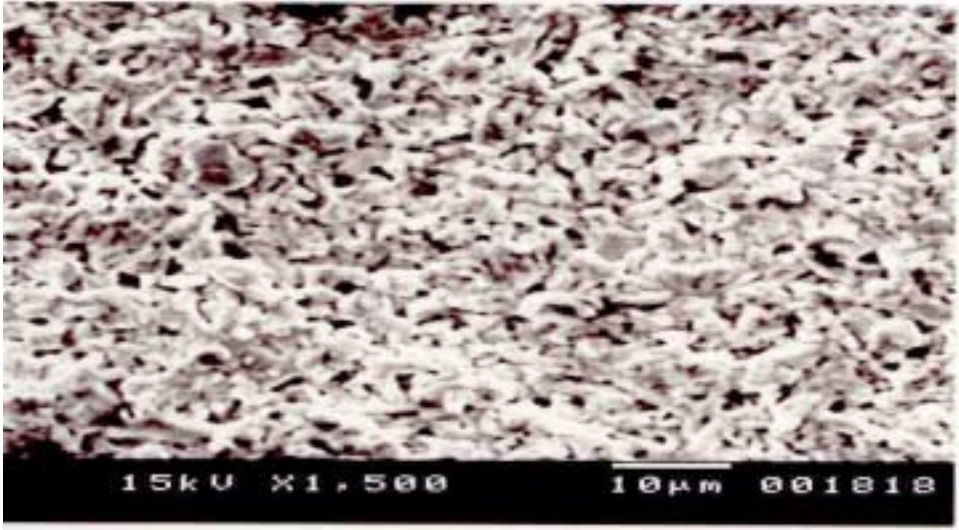

(a)

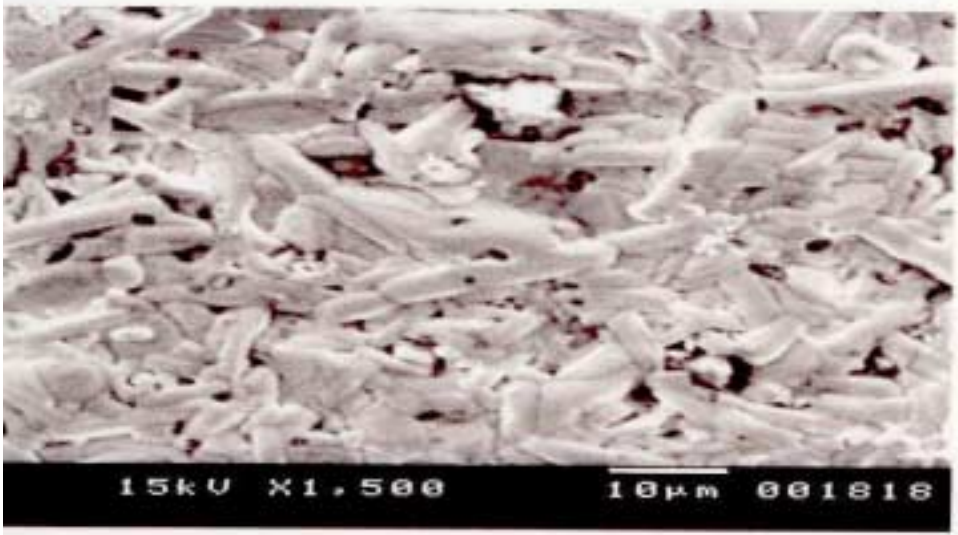

(b)

Fig. (3)

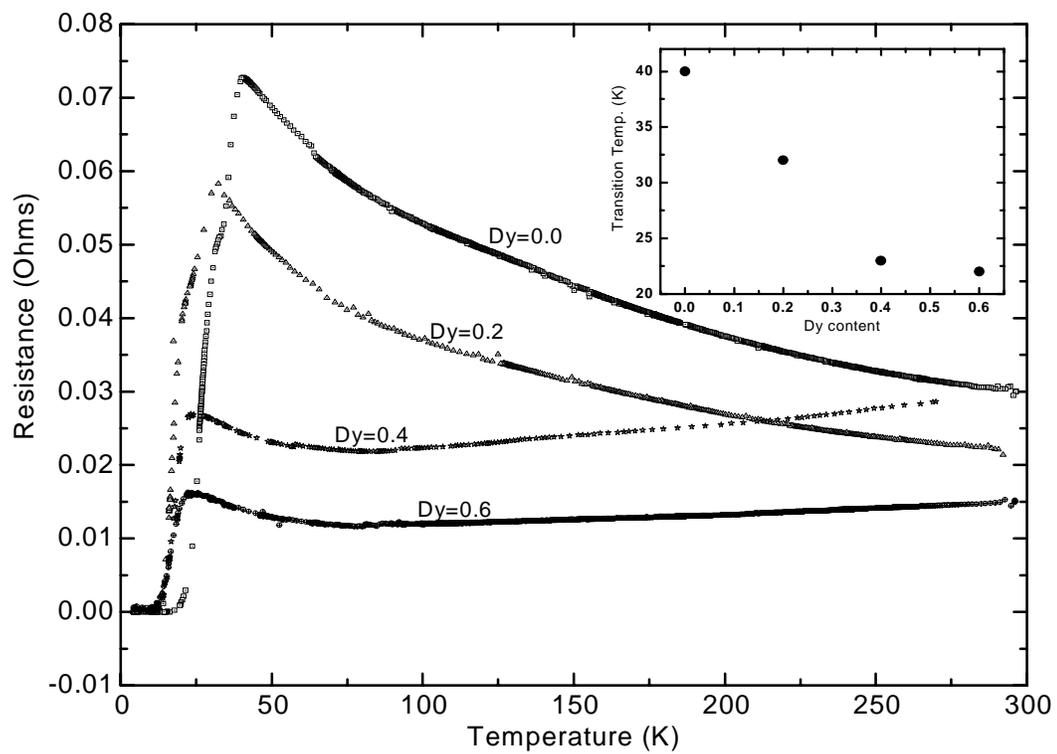

Fig. 4

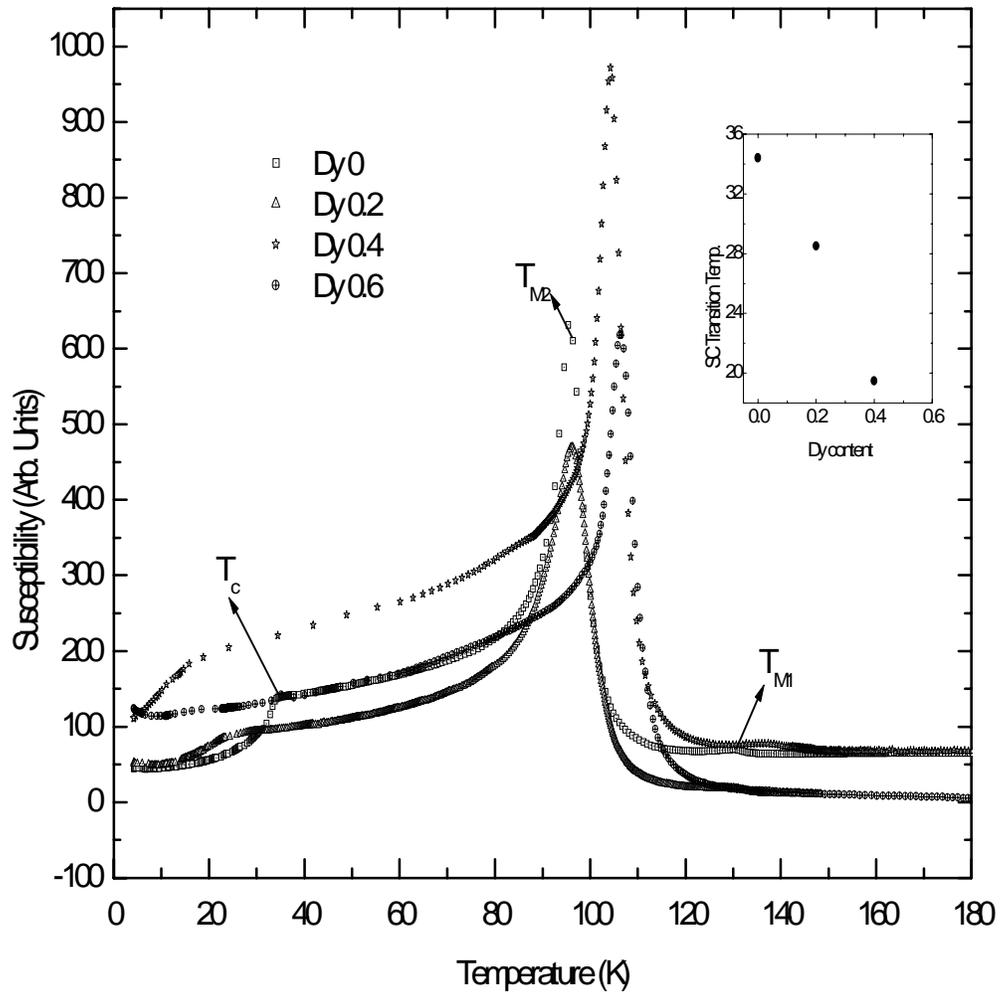

Fig. 5

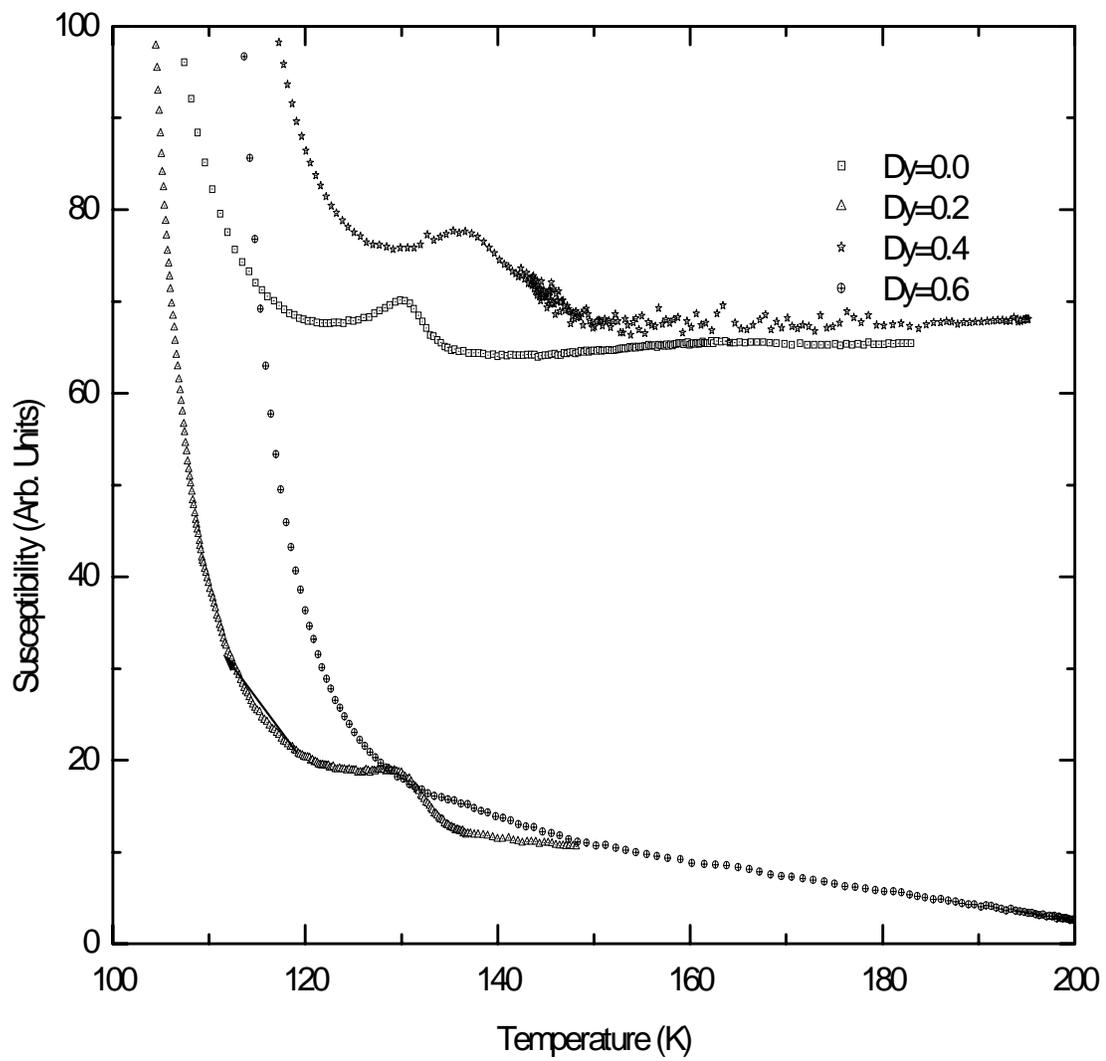

Fig. 6